\documentclass[journal]{IEEEtran}

\usepackage{amsfonts}
\usepackage{eucal}
\usepackage{amsmath}
\usepackage{amsthm}
\usepackage{fullpage}
\usepackage{cite}
\usepackage{graphicx}
\usepackage{multirow}
\usepackage{array}
\usepackage{url}
\usepackage{color}
\usepackage[linesnumbered,ruled]{algorithm2e}
\usepackage{amssymb,array,enumerate,epsfig,pifont,graphics,subfigure}

\begin{document}

\title{Hierarchical Model of Human Guidance Performance Based on Interaction Patterns in Behavior}

\author{B\'{e}r\'{e}nice Mettler and Zhaodan Kong
\thanks{This work was presented on the 2nd International Conference on Application and Theory of Automation in Command and Control Systems (ICARUS), London, UK, 2012.}
\thanks{The research has been made possible through the financial support from the National Science Foundation (Award No. NSF/CMMI-1002298).}
\thanks{Z. Kong is a postdoctoral associate with the Department of Mechanical Engineering, Boston University, 110 Cummington Mall, Boston, MA 02215 (email: zhaodan@bu.edu). This work was performed while he was a PhD student with the Department of Aerospace Engineering and Mechanics, University of Minnesota.\newline B. Mettler is an Associate Professor with the Department of Aerospace Engineering and Mechanics, University of Minnesota, 110 Union Street SE, Minneapolis, MN 55455 (corresponding author, email: mettler@umn.edu).}
}

\maketitle

\begin{abstract}
This paper describes a framework for the investigation and modeling of human spatial guidance behavior in complex environments. The model is derived from the concept of interaction patterns, which represent the invariances or symmetries inherent in the interactions between an agent and its environment. These patterns provide the basic elements needed for the formalization of spatial behavior and determine a natural hierarchy that can be unified under a hierarchical hidden Markov model. 
\end{abstract}

\section{Origin And Underlying Principles}

Spatial behavior has been an active research topic in psychology and robotics over the past few decades. What fascinates researchers is the ability of trained humans to spontaneously generate behavior for problems that are often not tractable from a computational standpoint~\cite{Mettler11}. Take driving a car for instance, it involves a driver, a car (the driver and the car together can be taken as the agent), as well as the surrounding environment. All three have their own dynamics. The driver needs not only to comprehend the dynamics of each single component, but also needs to have a holistic understanding of the dynamics of the entire agent-environment system. 

Furthermore, these types of problems generally involve processes that obey quite different principles. Sensing and perception are often considered to be probabilistic, while cognition and action are considered to be deterministic~\cite{campbell2010autonomous,mumford2002pattern}. A driver or pilot has to integrate all these processes while factoring in the overall goal of the task, e.g., driving to a specified location safely and as fast as possible.  


Theories regarding the organization of behavior can be categorized into two main schools: model-based approaches and non-representational approaches. Most non-representational approaches like tau coupling~\cite{lee1998guiding}, or more recently models based on information processing and dynamical principles~\cite{warren2006dynamics}, provide useful explanations of the perception-action loop in behavior. However, the behavior is almost always formulated in terms of some low-dimensional dynamics without specific meaning. Their main limitations is that they cannot explain complex behaviors involving the composition of several behaviors, such as those that result from more complex environments with a remote goal state. 

Model-based approaches~\cite{wolpert2000computational,todorov2004optimality}, on the other hand, tend to focus either on the perception or the action side; they are rarely presented under a unified framework. If they are, they are most often based on the generic ``sense-model-plan-act'' model, which due to its rigidity makes it challenging to explain the flexible and adaptive capabilities of human spatial behavior. 

This short paper only highlights the key concepts and results. For a comprehensive treatment of the concepts that we introduce, as well as the details regarding the experiments, the algorithms and the results, please refer to~\cite{kong2012pattern,kong2011foundations,kong2011investigation}.

\begin{figure}[!t]
\centering
\subfigure[]{\includegraphics[width=\columnwidth]{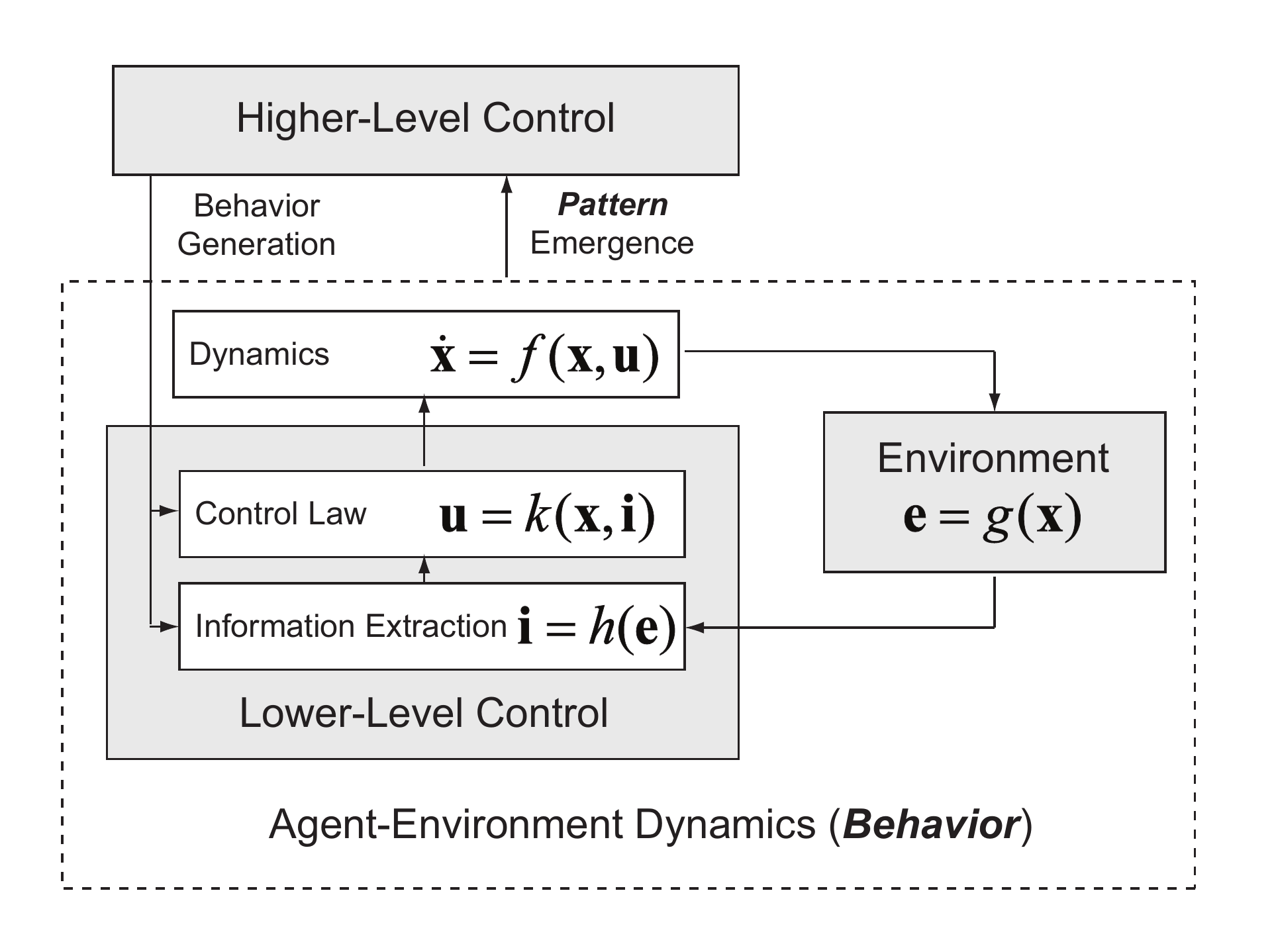}\label{sub:fig_two_layered_control_structure}}
\subfigure[]{\includegraphics[width=0.9\columnwidth]{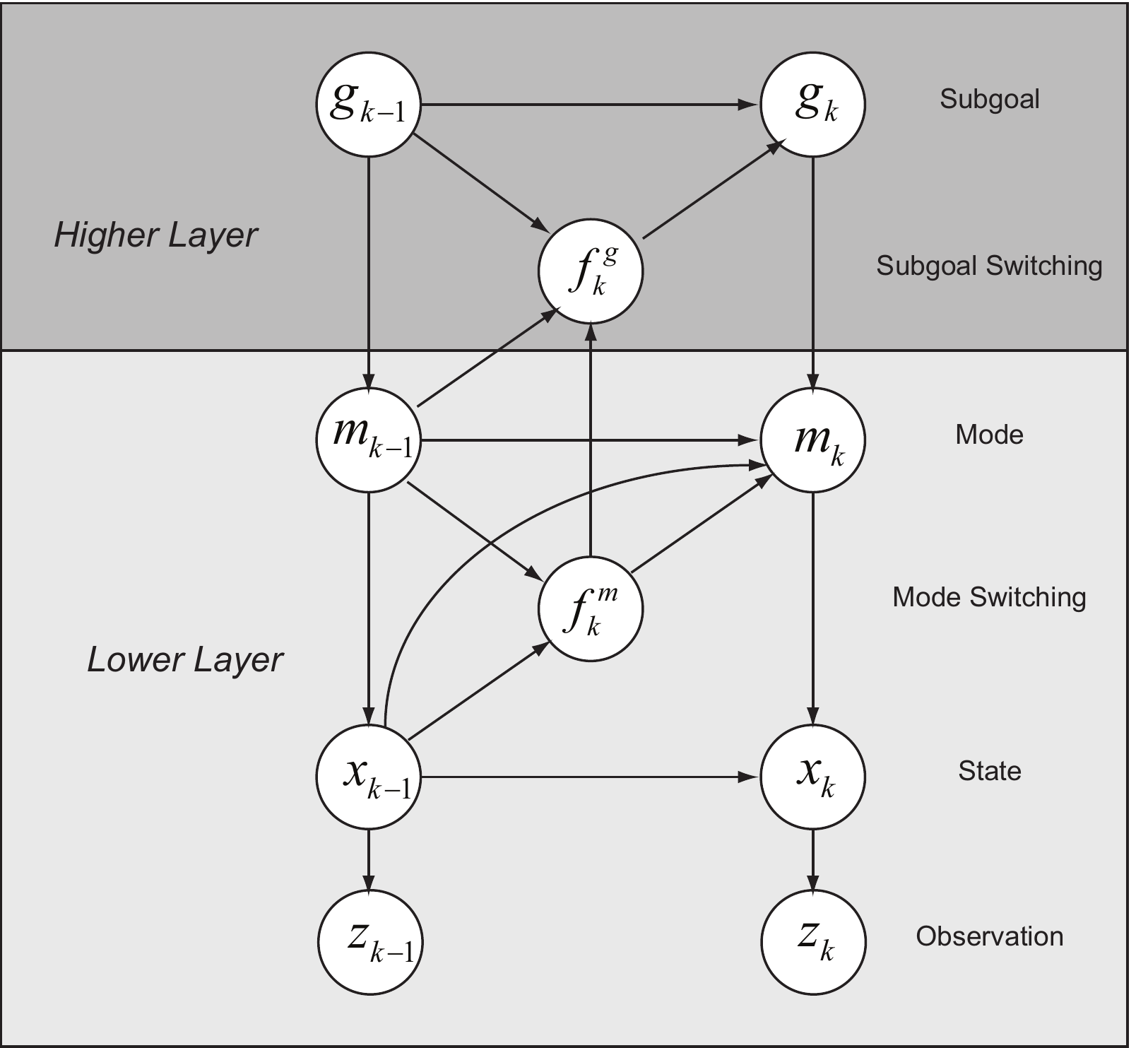}\label{sub:Hierarchical_Markov_Model}}
\caption{(a) A conceptual pattern-based hierarchical guidance model. (b) Hierarchical hidden Markov model of guidance behavior, comprising two layers with the interaction pattern serving at the link between the two.} \label{fig:Hierarchical_Markov_Model}
\end{figure}

\section{Modeled Relationships}


Considering the range of complexities involved in the agent-environment dynamics and the perception-cognition-action processes, one of the fundamental question that need to be addressed in the study of spatial behavior is what aspects of these dynamics are fundamental in explaining how spatial behavior is organized.  Our modeling framework is built on the analysis of the agent-environment dynamics  as illustrated in Fig.~\ref{sub:fig_two_layered_control_structure} and focuses on the understanding and characterizing the emerging interaction patterns, and how these can then help formalize the analysis of behavior and the development of a model like the HHMM shown in Fig.\ref{sub:Hierarchical_Markov_Model}.

\subsection{Agent-Environment Dynamics and Interaction Patterns}

\begin{figure}[!t]
\centering   \includegraphics[width=\columnwidth]{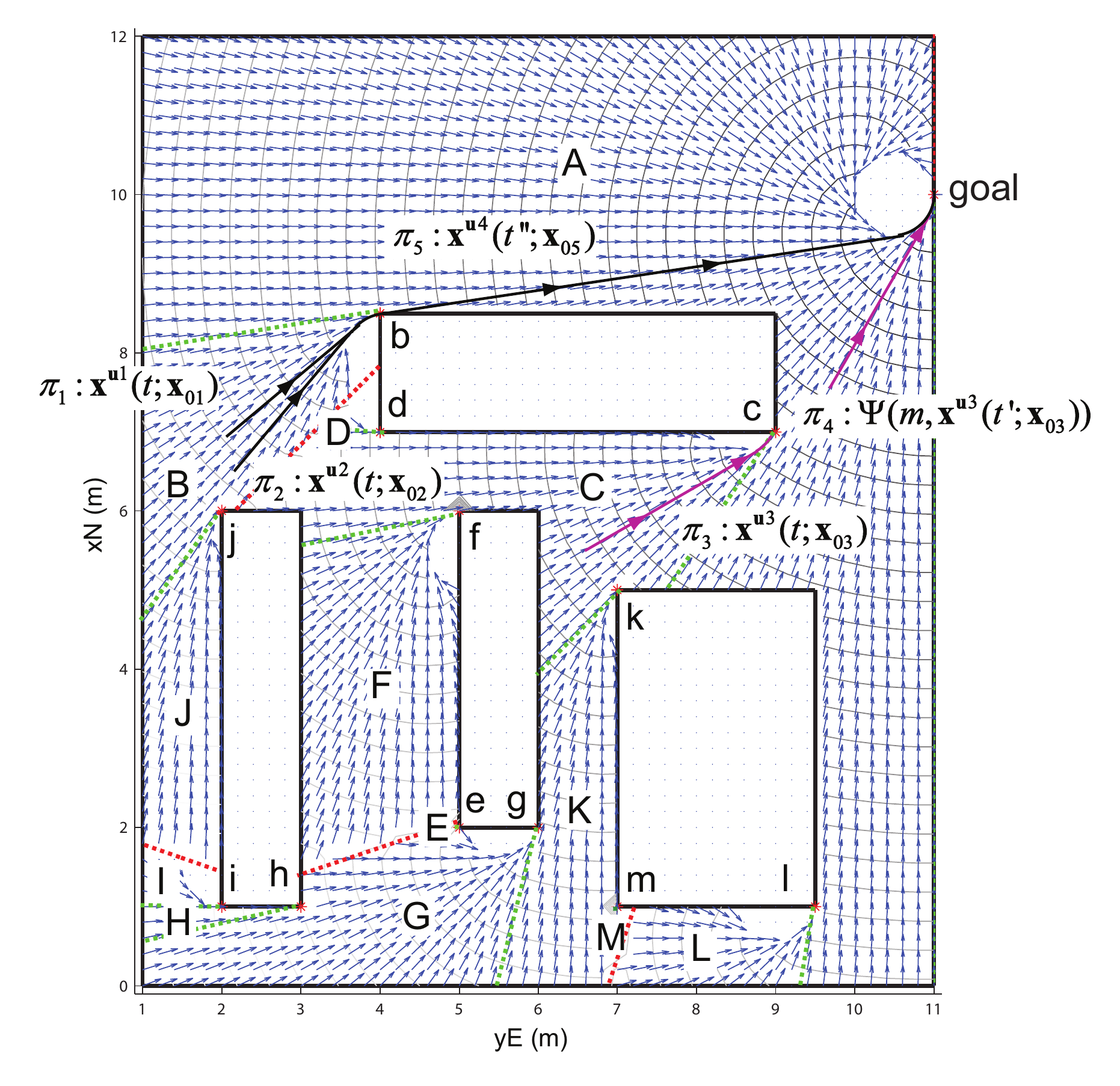}
    \caption{Optimal guidance behavior of Dubins' vehicle for an environment with multiple obstacles.}
    \label{f:Dubins_Behavior}
\end{figure}


The dynamics of an agent can be described as:
\begin{equation}
\dot{\mathbf{x}} = f(\mathbf{x},\mathbf{u}) \label{e:agent_dynamics}
\end{equation}
where $\mathbf{x} \in \mathcal{X} \subseteq \mathbb{R}^n$ is the agent state, $\mathbf{u} \in \mathcal{U} \subseteq \mathbb{R}^m$ is the control law. In order for an agent to perform a guidance task, it should be able to perceive the world via $\mathbf{i}(t) = h(\mathbf{e}(t))$, where $\mathbf{i}(t)$ is the information and $\mathbf{e} \in \mathcal{E}$ is the environment state, which can be represented as $\mathbf{e}(t) = g(\mathbf{x}(t))$, and choose an appropriate action $\mathbf{u}(t)$ to bring itself from an initial state $\mathbf{x}_0$ to a goal state $\mathbf{x}_f$. The process of choosing the right action can be written in the form of a control policy as $\mathbf{u}(t) = k(\mathbf{x}(t),\mathbf{i}(t))$. Putting all these components together results in the following closed-loop agent-environment dynamics:
\begin{equation}
\dot{\mathbf{x}} = f(\mathbf{x},k(\mathbf{x},h(g(\mathbf{x}))))
\label{e:behavior_definition}
\end{equation} 
The collection of all the agent state trajectory, $\{\mathbf{x}(t): t \in [t_0,t_f]\}$, together with the corresponding environment state trajectory, $\{\mathbf{e}(t): t \in [t_0,t_f]\}$, with respect to this closed-loop dynamics is our formal definition of \textit{guidance behavior}~\cite{kong2011foundations}. 

We can introduce two types of relationships over guidance behavior. One relationship $\sim_g$ is defined by extending the concept of motion primitive~\cite{frazzoli2005maneuver}: two trajectories, $\overleftarrow{s_i}^L$ and $\overleftarrow{s_j}^L$\footnote{For computational convenience, both the $\sim_g$ and $\sim_s$ relationships are defined on the symbolic representations of guidance behavior instead of the continuous form (\ref{e:behavior_definition}), which can be obtained through a quantization, $q: \mathcal{X} \times \mathcal{E}  \rightarrow \mathcal{A}$, where $\mathcal{A}$ is a finite set of symbols which is called the state alphabet. With such a quantization, the set of all trajectories can then be written as: $\overleftarrow{S} = \{\overleftarrow{s}^L_i:s_{i-L+1},...,s_{i}, s \in
\mathcal {A}, L \in \mathbb{Z}^+, i \in \mathbb{Z}\}$. The controls can be quantized similarly.}, satisfy $\overleftarrow{s_i}^L \sim_g \overleftarrow{s_j}^L$ if there exist a $m \in M$ and control histories, $\overleftarrow{u}^L_i$ and $\overleftarrow{u}^L_j$ such that:
\begin{equation}
(\Psi(m,\overleftarrow{s}^L_i),\overleftarrow{u}^L_i) = (\overleftarrow{s}^L_j,\overleftarrow{u}^L_j)
\label{e:equivalence_relationship_definition_1}
\end{equation} 
where $L\in \mathbb{Z}^+$, $M$ is some finite-dimensional Lie group, and $\Psi$ is the action of this group $M$ on the state manifold $\mathcal{X}$, $\Psi:= M \times \mathcal{X} \rightarrow \mathcal{X}$. The following two conditions need to be held for all $m \in M$, $\mathbf{x} \in \mathcal{X}$, $\mathbf{e} \in \mathcal{E}$, $t \in [t_0,t_f]$ and all $u \in \mathcal{U}$ in order to satisfies (\ref{e:equivalence_relationship_definition_1}): 
\begin{equation}
\Psi(m,\mathbf{x}^\mathbf{u}(t;\mathbf{x}_0)) = \mathbf{x}^\mathbf{u}(t;\Psi(m,\mathbf{x}_0))
\label{e:symmetry_definition}
\end{equation} 
and

\begin{equation}
\Psi|\mathcal{E}(m,\mathbf{e}^\mathbf{u}(t;\mathbf{e}_0)) = \mathbf{e}^\mathbf{u}(t;\Psi|\mathcal{E}(m,\mathbf{e}_0))
\label{e:symmetry_environment_definition}
\end{equation} 
where $\Psi|\mathcal{E}$ is the restriction of mapping $\Psi$ in $\mathcal{E}$ and it can be well defined on the assumption that the environment state can be written in the form of some relative quantities, $\mathbf{e}(t) = \mathbf{x}(t) - \mathbf{x}_r$. (\ref{e:symmetry_definition}) implies that if $t \rightarrow (\mathbf{x}(t),\mathbf{u}(t))$ is an integral curve of (\ref{e:agent_dynamics}), so is $t \rightarrow (\Psi(m,\mathbf{x}(t)),\mathbf{u}(t))$. (\ref{e:symmetry_environment_definition}) can be interpreted similarly. 

The other relationship $\sim_{s}$ is defined using the concept of causal state~\cite{shalizi2001computational}: 
\begin{equation}
\overleftarrow{s_i}^K \sim_{s} \overleftarrow{s_j}^L
\Leftrightarrow P(\overrightarrow{S}|\overleftarrow{s_i}^K) =
P(\overrightarrow{S}|\overleftarrow{s_j}^L)
\label{e:equivalence_definition}
\end{equation}
for all semi-infinite $\overrightarrow{S} = s_0 s_1...$, where $K,L
\in \mathbb{Z}^+$ and $P$ stands for probability. Since, in this paper, we are only concerned with deterministic systems, we can then assign $P$ equal to one. In this setting, the state $s_0$ is then called a \textit{subgoal}, in the sense that, from this state on, the two trajectories, $\overleftarrow{s_i}^K$ and $\overleftarrow{s_j}^L$, will follow the same trajectories $\overrightarrow{S}$. We will drop the length variables $K$ and $L$ here and denote the members of any length in the set $\overleftarrow{S}$ by $\overleftarrow{s}$. 

It can be easily verified that both $\sim_g$ and $\sim_s$ are equivalence relationships. Thus, if $\overleftarrow{s} \in \overleftarrow{S}$, one type of equivalence class over $\overleftarrow{S}$ can be defined in the following two steps (the order of these two operations is interchangeable): 
$[\overleftarrow{s}] = \{ \overleftarrow{s}' \in \overleftarrow{S}:
\overleftarrow{s}' \sim_s \overleftarrow{s}\}$
and 
$[[\overleftarrow{s}]] = \{ [\overleftarrow{s}]' \in [\overleftarrow{s}]:
[\overleftarrow{s}]' \sim_{g} [\overleftarrow{s}]\}
$. Each equivalence class $[[\overleftarrow{s}]]$ is called an \textit{interaction pattern} to reflect the fact it captures invariances inherent in the agent-environment interactions.

For the guidance behavior of Dubins' vehicle as shown in Fig.~\ref{f:Dubins_Behavior}, the agent state and the environment state are invariant with respect to a translation and a rotation about a vertical axis, or to the actions of the symmetry group $M = SE(2)$. Each element of $M$ can be written in the form of a 3$\times$3 matrix $m(\psi,\mathbf{t})$, with rotation angle $\psi$ and the translation vector $\mathbf{t} = [t_x,t_y]'$. For the example trajectories shown in Fig.~\ref{f:Dubins_Behavior}, according to (\ref{e:equivalence_relationship_definition_1}) and (\ref{e:equivalence_definition}), we have $\pi_1 \sim_s \pi_2$ and $\pi_3 \sim_g \pi_4$. And taking the $\sim_s$ equivalence (e.g., equating the two black trajectories) results in the partitions of the state space (encircled by obstacle boundaries, red and green lines, which correspond to repelling and attracting manifolds, respectively~\cite{kong2011investigation}), and subsequently taking the $\sim_g$ equivalence (e.g., equating the two purple trajectories) results in two interaction patterns: one corresponds to approaching the subgoals from the left, such as guidance behaviors in partition A, and the other corresponds to approaching the subgoals from the right, such as guidance behaviors in partition B. Actually, after a mirror reflection symmetry is added to group $M$, only a single interaction pattern is left.

\subsection{Experimental Investigation and Validation}

Next we proceeded to investigate the agent-environment dynamics and validate the concept of interaction pattern using experimental data.  The approach together with the necessary computational tools from machine learning, system identification and pattern recognition are summarized in the following five steps (I-V). 

\begin{figure}[!t]
\centering
\subfigure[]{\includegraphics[width=0.8\columnwidth]{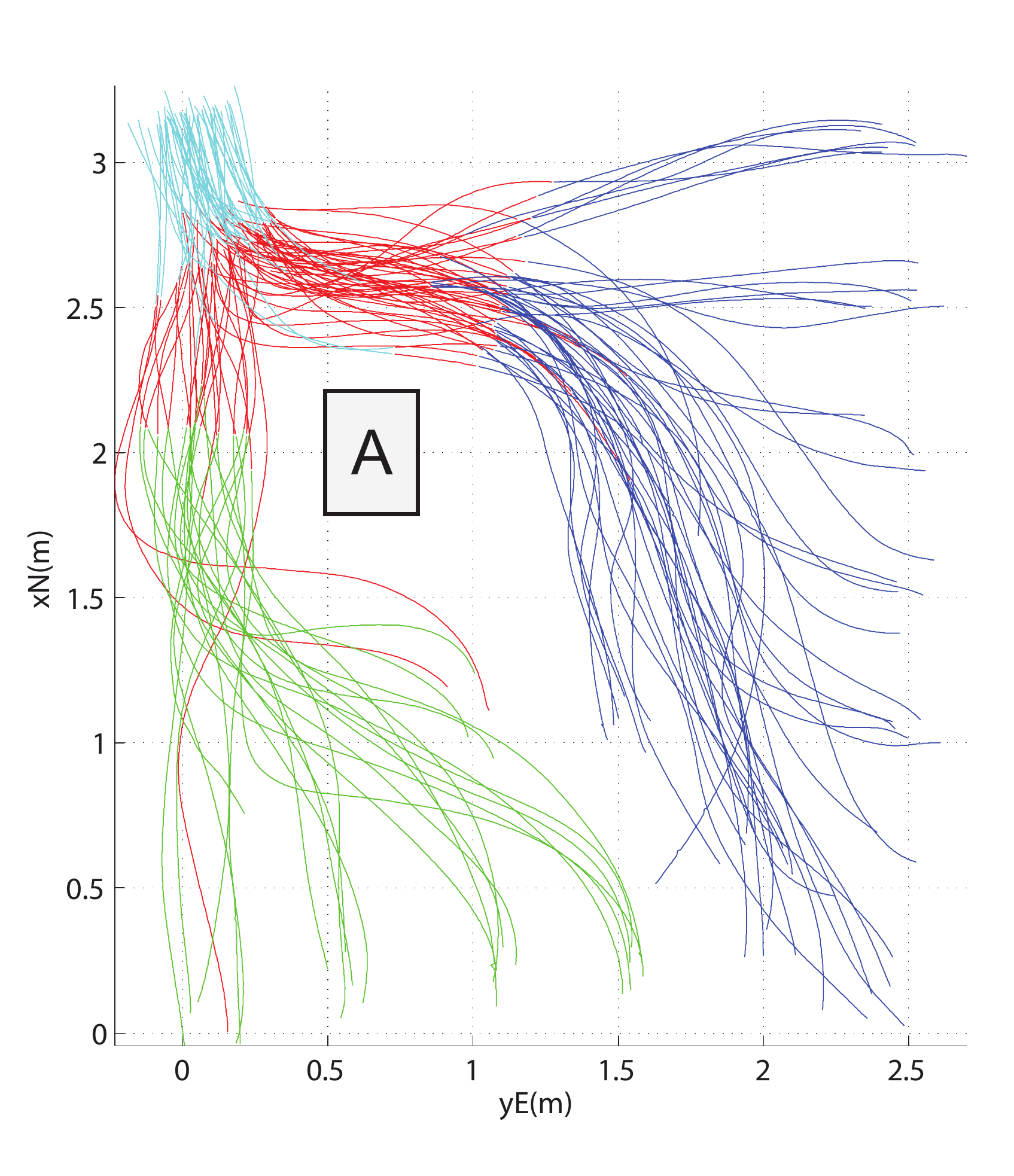}\label{sub:Trajectory_Segmentation_A}}
\subfigure[]{\includegraphics[width=0.8\columnwidth]{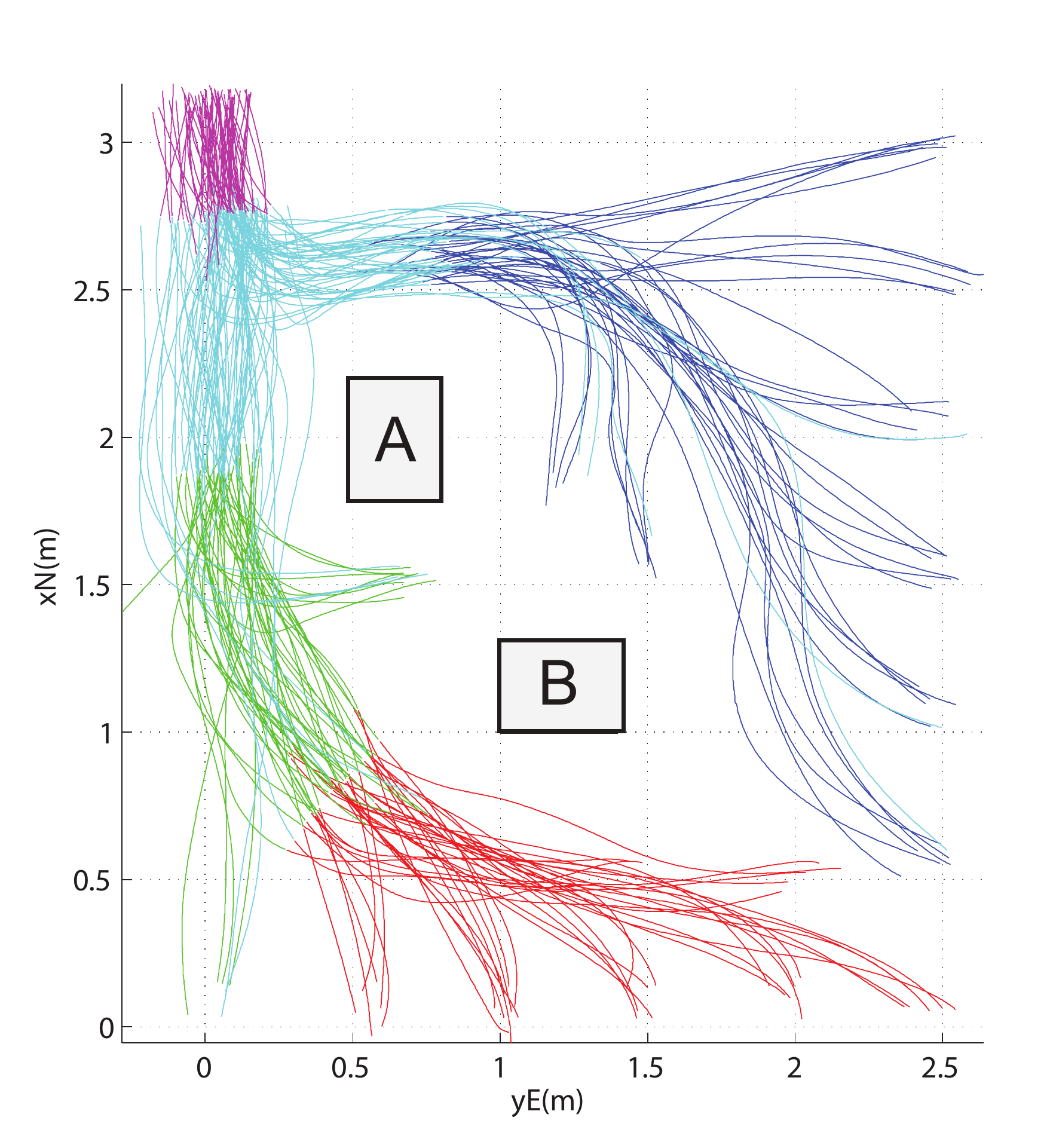}\label{sub:Trajectory_Segmentation_C}}
\caption{Trajectory segmentation results for two different tasks.} \label{fig:clustering_total}
\end{figure}

Experimental trajectory data was collected from agile guidance tasks performed with a miniature remote control helicopter~\cite{kong2011investigation} (Fig.~\ref{fig:clustering_total}) in our interactive guidance and control lab~\cite{mettler2012lab}. The data was represented as $\mathbf{x}^n(i), i = 1,...N_n, n = 1,...,N$ with $N$ as the number of trajectories and $N_n$ as the number of data point for trajectory $n$. The helicopter planar rigid-body motion is fully characterized by four variables $\mathbf{x}^n(i):=[x^n(i),y^n(i),v^n(i),\psi^n(i)]'$, where $[x^n(i),y^n(i)]'$ is the position, $v^n(i)$ is the speed and $\psi^n(i)$ is the course angle. 

\underline{(I) Symbolic representation and subgoal identification}: Transformation of the trajectory data into a symbolic representation and identifying pairwise subgoals based on the definition (\ref{e:equivalence_definition}). The transformation is done by quantizing the state space into mutually exclusive cells according to $q$. Each cell is a letter of the state alphabet $\mathcal{A}$. If a data point $\mathbf{x}^n(i)$ falls within a cell, it is represented by the corresponding letter $s^n(i)$. Once the transformation is done for each data point, the original measurement data $\mathbf{x}^n(i), i = 1,...N_n, n = 1,...,N$ is transformed into its corresponding symbolic form as $s^n(i), i = 1,...N_n, n = 1,...,N$.

\underline{(II) Subgoal clustering and trajectory segmentation}: The extracted subgoals are clustered applying Isomap, multidimensional scaling and K-means methods. The experimental trajectories are then clustered into $\sim_s$ equivalent segment clusters based on their membership subgoal.\footnote{The application of our clustering operation is based on the assumption that the ``observed'' subgoals extracted from step (I) are the expression of some ``hidden'' subgoals, where the number of hidden subgoals is much smaller than the observed ones. The distribution of the observed subgoals can be modeled by a mixture of Gaussians as follows~\cite{bishop2006pattern}.} The original trajectory sample data are augmented by their memberships as follows:
\begin{equation*} 
[\mathbf{x}^m(i)',\xi^m(i)]', i = 1,...M_m, m = 1,...,M
\end{equation*} 
with $M$ as the number of trajectory segments, $M_m$ as the number of data point for trajectory segment $m$, and $\xi^m(i)$ as the membership of data point $\mathbf{x}^m(i)$ with $1 \leq \xi^m(i) \leq K^*$. The segmentation results for two tasks are shown in Fig.~\ref{fig:clustering_total}. 

\begin{figure}[!t]
\centering
\subfigure[]{\includegraphics[width=0.9\columnwidth]{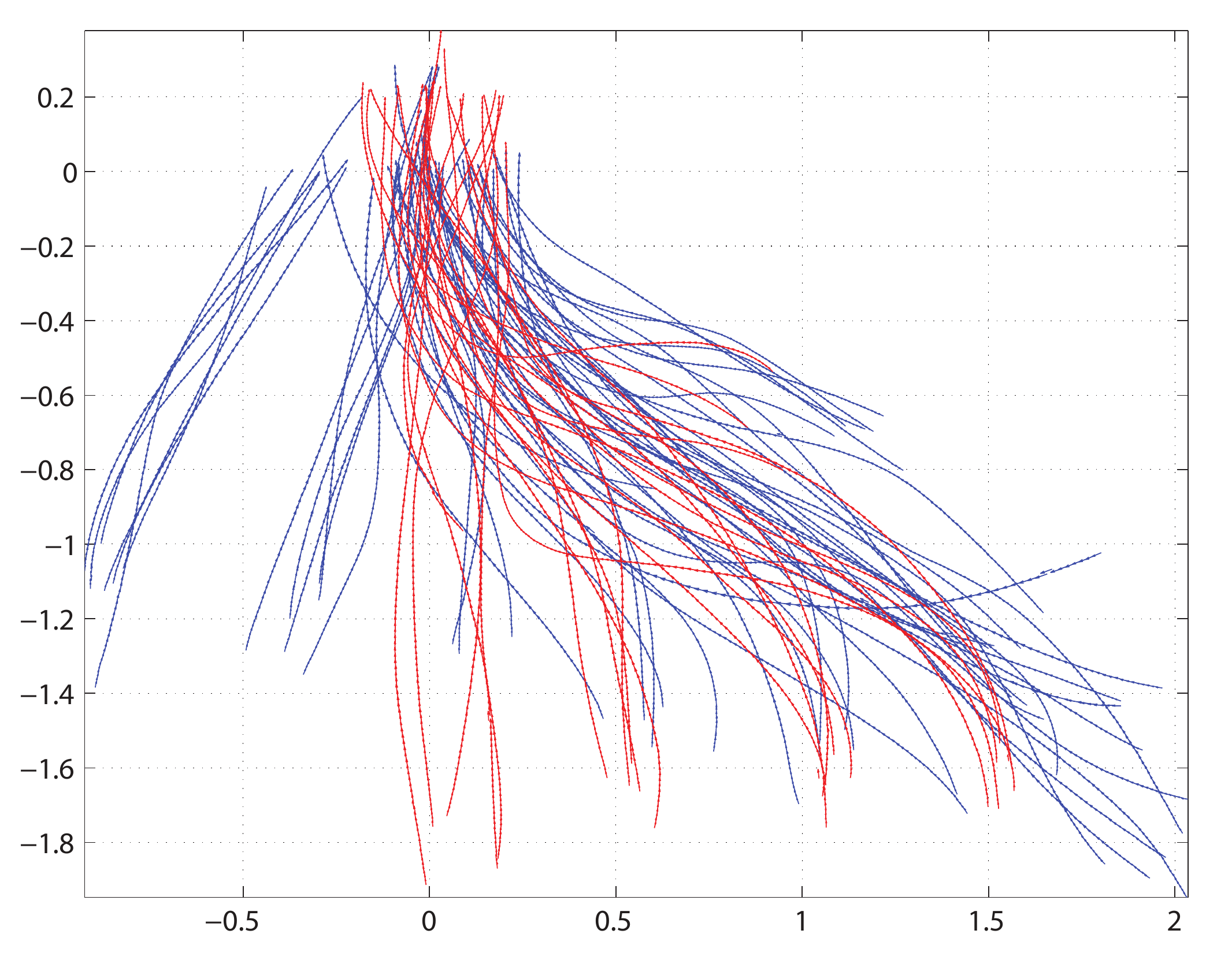}\label{sub:Matching_TJ}}
\subfigure[]{\includegraphics[width=0.9\columnwidth]{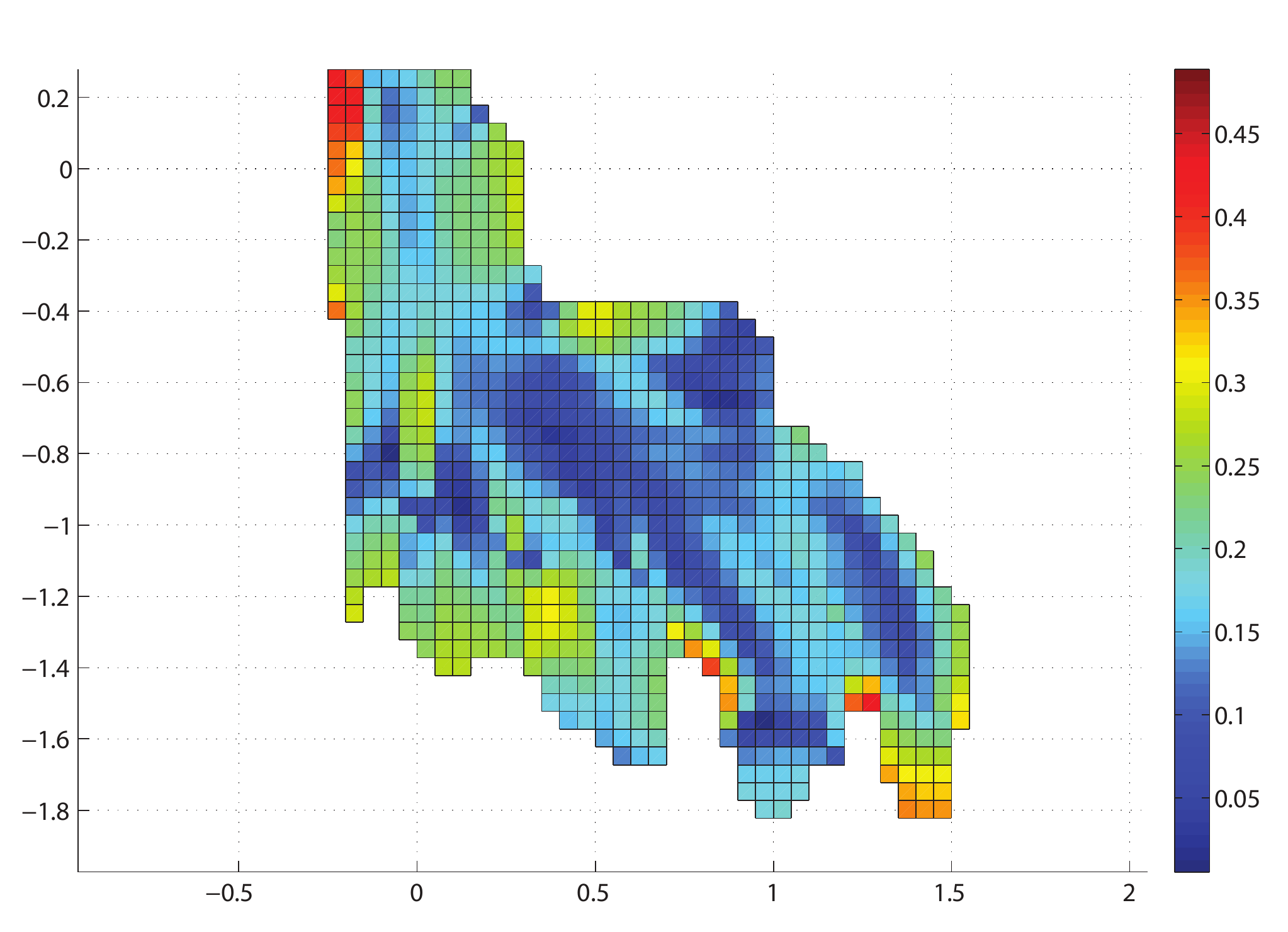}\label{sub:Matching_RD}}
\caption{Matching results. (a) Superimposed trajectory segments. (b) The relative difference between correspondence points.} \label{fig:Single_Matching}
\end{figure}

\underline{(III) $\sim_g$ equivalence analysis}: To prove that one segment cluster $\xi_1$ is symmetric to another cluster $\xi_2$ according to $\sim_g$ equivalence definition (\ref{e:equivalence_relationship_definition_1}) or they both belong to a same interaction pattern, it suffices to show that for any trajectory $\mathbf{x}^u(t;\mathbf{x}_0)$ from cluster $\xi_1$, there exists an action $m$ from the symmetry group $M$ and a trajectory $\mathbf{x}^{u'}(t';\mathbf{x}'_0)$ from cluster $\xi_2$ such that $\mathbf{x}^{u}(t';\mathbf{x}'_0) = \Psi(m,\mathbf{x}^u(t;\mathbf{x}_0))$ holds and all the m's are the same, and vice versa. This evaluation was formulated as a parameter optimization problem. The matching results for two of the clusters are shown in Fig.~\ref{fig:Single_Matching}.

\begin{figure}[!t]
\centering
\subfigure[]{\includegraphics[width=0.8\columnwidth]{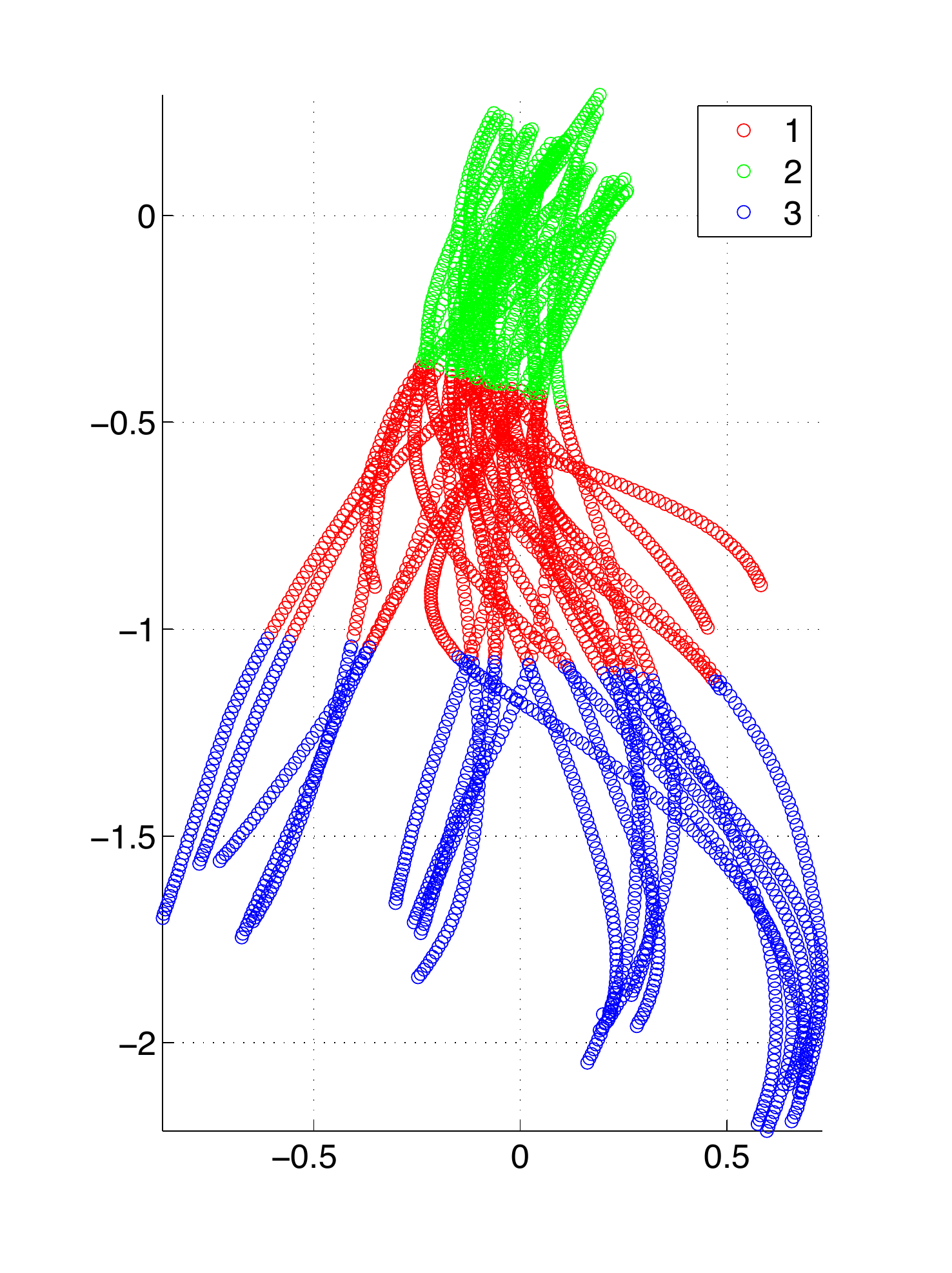}\label{sub:PWA_After_Discriminant}}
\subfigure[]{\includegraphics[width=\columnwidth]{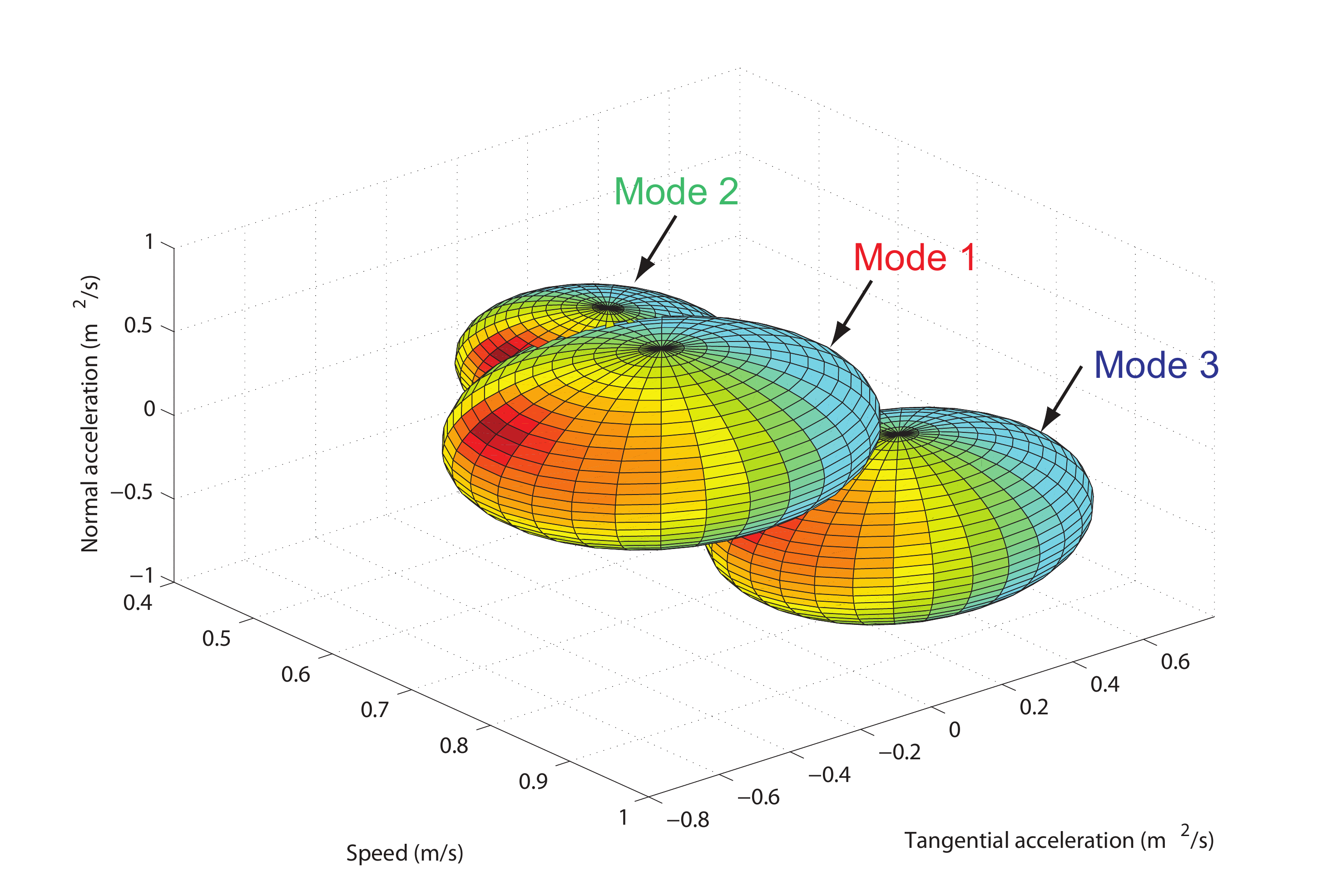}\label{sub:Ellipsoid_Pilot_A}}
\caption{(a) Data points according to their mode memberships. (b) shows a typical fitted distribution of velocity, tangential and normal accelerations for a segment cluster.} \label{fig:PWA_model}
\end{figure}

\underline{(IV) Analysis of dynamical characteristics}: The dynamical characteristics of guidance behavior in each segment clusters (as sample interaction patterns) is analyzed using piecewise affine (PWA) model~\cite{ferrari2003clustering}. A PWA system is defined by the following state-space equations:
\begin{equation}
\mathbf{x}(t+1)=A_i\mathbf{x}(t)+B_i\mathbf{u}(t)+d_i, \text{ for } \mathbf{x} \in \mathcal{X}_i, \mathbf{u} \in \mathcal{U}_i
\label{e:PWA_model}
\end{equation}
where $\{\mathcal{X}_i\}_{i=1}^{l_m}$ and $\{\mathcal{U}_i\}_{i=1}^{l_m}$ are polyhedral partitions of $\mathcal{X}$ and $\mathcal{U}$ (each partition can be called a mode), and $d_i$ is the noise term. As shown in Fig.~\ref{fig:PWA_model}, for each segment cluster, three modes with distinguished characteristics can be identified. They are a starting mode (mode 3) $m^s$, a coasting mode (mode 1) $m^c$, and an approaching mode (mode 2) $m^a$.

\begin{figure}[!t]
\centering
\subfigure[]{\includegraphics[width=0.8\columnwidth]{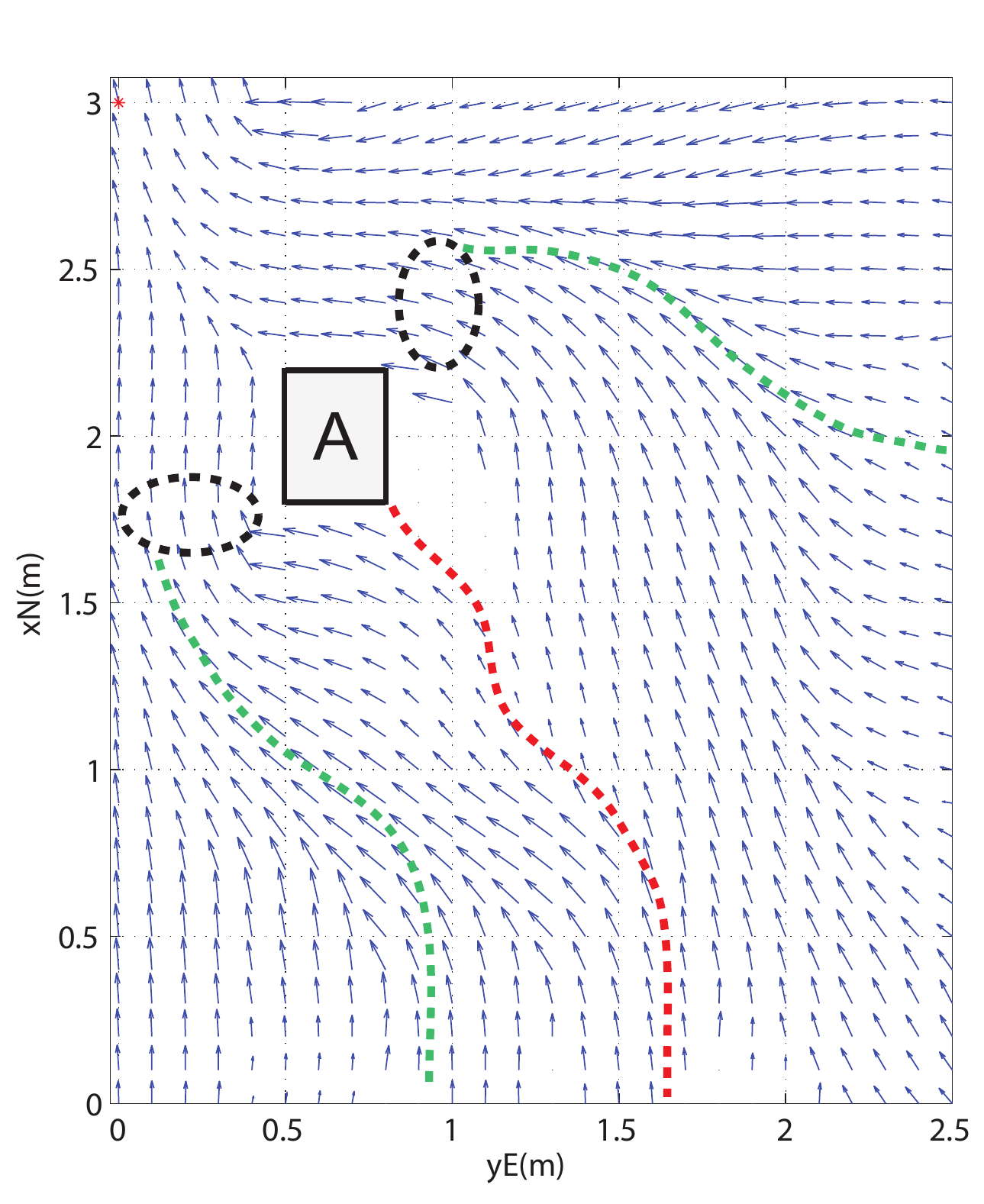}\label{sub:VF_A}}
\subfigure[]{\includegraphics[width=\columnwidth]{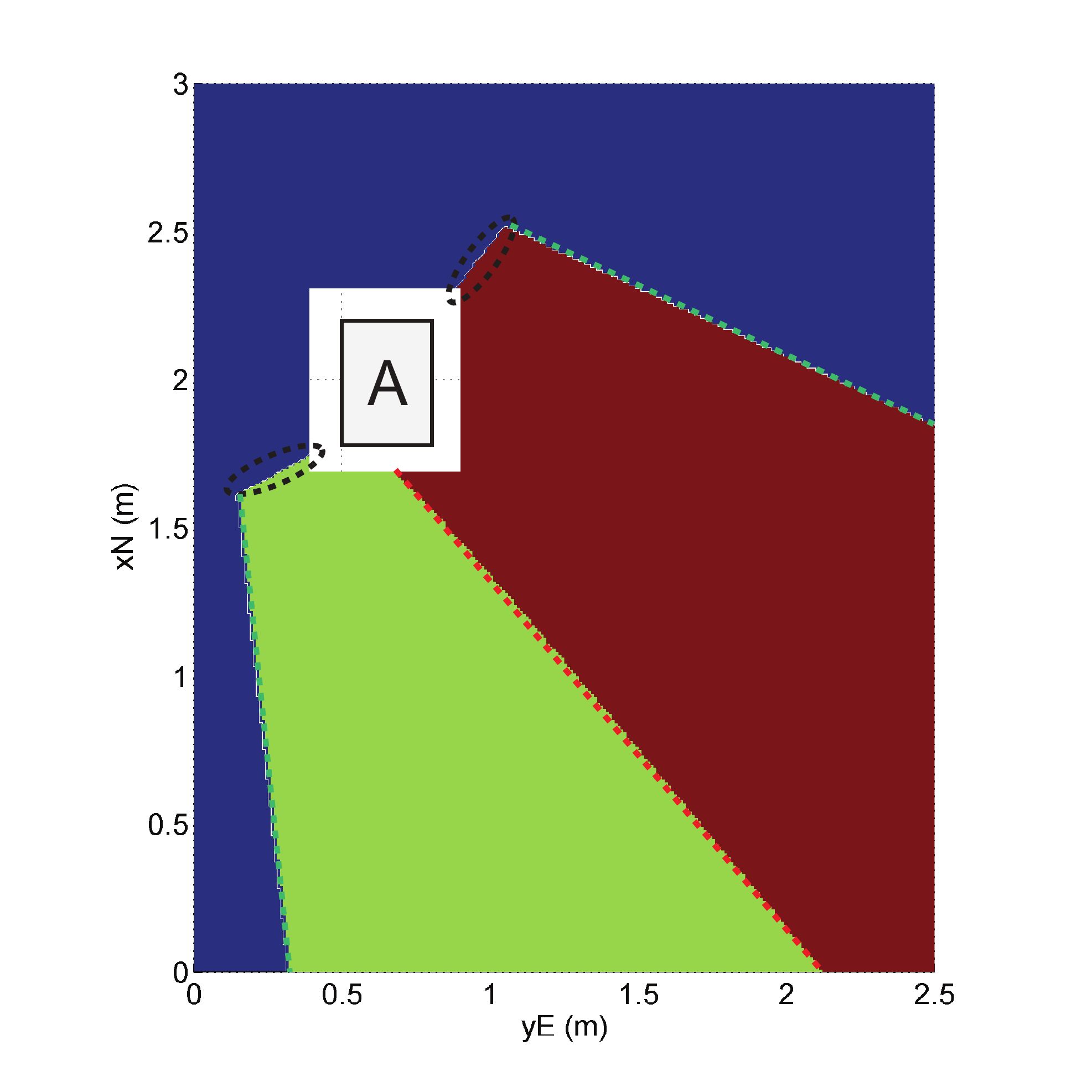}\label{sub:Prediction_Membership_A}}
\caption{(a) Vector field computed from experimental data. (b) Predicted partition. In both of these two figures, black lines mark the locations of subgoals, red line marks the location of repelling manifold, and green lines mark the locations of attracting manifolds.} \label{fig:meta_analysis}
\end{figure}

\underline{(V) Meta-behavioral analysis}: The transition among the segment clusters and their spatial boundaries are investigated based on general dynamical systems concepts. The transition boundaries among the patterns can be approximately characterized as attracting and repelling manifolds~\cite{kong2011foundations,kong2011investigation}. The functional form of time-to-go (TTG) function is first learned from experimental data and then a wavefront method based on optimality principles can be used to derive the partitions: the subgoals are determined as the locations where the wavefront defined by the learned TTG function meet the vertices of obstacles; the repelling manifolds correspond to the locations where two wavefronts originating from either a goal or a subgoal meet; and the attracting manifolds correspond to subgoals and their directions are determined by the gradient of the wavefront. Fig.~\ref{fig:meta_analysis} shows the predicted partitions that result from this method compared to the original partitions. 

\subsection{Integration under a Hierarchical Hidden Markov Model}

In summary, steps I and II correspond to taking the $\sim_s$ equivalence. Each extracted segment cluster can be seen as a sample interaction pattern. In step III, the $\sim_g$ equivalence of these segment clusters is evaluated. The results from the first three steps confirm that interaction patterns do exist in human guidance behavior and that they can be explained using equivalence concepts. Following, in step IV and V, the micro and macro organizational principles within and across these patterns are investigated. Here the results show that the transitions between modes within and across the patterns can be described through simple rules.

The analysis of guidance behavior based on interaction patterns suggests that guidance behavior follows a natural and systematic hierarchical organization. The overall system can be formalized using a hierarchical hidden Markov model (HHMM) as shown in Fig.~\ref{sub:Hierarchical_Markov_Model}. For the example in this paper, state $\mathbf{x}$ is taken as $[x,y,v,\psi]'$ and the measurement is taken to be the same as $\mathbf{x}$. Mode $m$ can take on three values: $m^s$, $m^c$ and $m^e$. The edges or dependencies among $m$ and $x$ at different times, along with the Boolean mode switching node $f^m$, are learned in step IV. Together, they correspond to the PWA systems learned from the interaction patterns.  Similarly, the edges among subgoals $g$ and the Boolean goal switching mode $f^g$ can also be learned from experimental data in step V.  

\section{Problem Addressed}

Our method based on an analysis of the guidance behavior in terms of agent-environment dynamics enabled to identify that the keystone in the organization of spatial behavior represents the invariances inherent to guidance behavior. These were described as interaction patterns and then used to formalize the guidance behavior under a hierarchical HHMM.  

Similar efforts of building formal models to study human behavior can also be found in vision~\cite{mumford2002pattern} and motor control~\cite{wolpert2000computational}. Our framework distinguishes itself by encompassing the entire perception-cognition-action loop. Furthermore, compared to some non-representational frameworks~\cite{lee1998guiding,warren2006dynamics}, thanks to the hybrid nature of our model, our framework can be easily extended and generalized to investigate more complex scenarios and behaviors.

Our framework also provides an avenue for understanding the organizational mechanisms humans and animals may utilize in order to reduce the burden of planning as well as to enable flexible and adaptive behavior. Our model suggests that high-level planning can be performed using an interaction pattern library, which can be understood as the repertoire of guidance capability that accounts for the agent-environment interactions. The cardinality of this library is much smaller than that of the entire state space. The results also show that explicit details of the agent's dynamics are not necessary for planning; it is how those dynamics manifest in the interaction patterns that really matter. 

The HHMM model shows that once a composition of interaction patterns has been elaborated, the pilot must primarily monitor whether the subgoal corresponding to the currently employed interaction pattern is attained and whether the interaction pattern remains valid. As long as the goal is not attained, the same subgoal and information extraction law $h(.)$ and control law $k(.,.)$ are applied. Once it is, a new subgoal, information extraction law $h(.)$ and control law $k(.,.)$ are initiated. 


Finally, it is important to underscore, that our framework relies largely on a data-driven approach. Assumptions regarding the nature and mechanisms underlying guidance behavior are kept to a minimum; the knowledge used to build the key elements of our model is almost entirely derived from the invariances that exist in the interactions between the agent and the environment. The details about the functional form of our interaction patterns, the number of them, the laws dictating the transitions between one pattern to the next could in principle all be learned from experimental data.

\section{Applications}

The HHMM model provides both descriptive and predictive capacities. This makes it useful for a range of engineering and scientific applications.  Being able to predict the pilot's behavior and performance based on the environment, task and mission elements is relevant to a number of applications. The model could be used as part of an active cueing system. Predicting behavior allows to identify potential failure states and then alerting the pilot and/or switching control modality.  

The hierarchic model delineates the relevant functions and levels of representation.  This knowledge can be used to determine the different modalities of interactions available to an operator and will help determine the design specifications for a broader range of human-machine control modalities. 

Another application for our framework is the development of novel planning and control algorithms for autonomous systems. For instance, one ongoing challenge in robotics is the brittleness of robot's performance~\cite{campbell2010autonomous}. The gained knowledge of the principles that dictate the organization of spatial behavior will support our understanding of the adaptive guidance capabilities and in turn help design algorithms needed to operate in less structured and partially known environments.

\section{Limitations And Development Opportunities}

Due to the limited number of scenarios, only the functional form of the time-to-go are learned while the organizational rule is assumed to follow some optimality principle. Although the prediction of the locations of subgoals and boundaries are reasonable, data from a broader range of experiments are needed to learn the high-level transition laws from experimental data.   

In terms of development opportunities, the modeling framework provides a new way to study perceptual and control mechanisms. The organization of behavior based on the interaction patterns must be intimately linked to the perceptual mechanisms. In fact these patterns are the manifestation of the perception-action loop. We are currently conducting experiments with eye tracking device determine relationships between attention patterns and behavior. These experiments will help us account for the specific perceptual mechanisms in the agent-environment model.   

Following the same vein, the functional description provided by the model makes it possible to understand what potential measurements can be used to investigate operator workload and attention. Brain imaging and brain activity analysis is still often treating the brain as a black box. Our hierarchic model provides a more precise picture of the type of activation levels (control, perceptual, planning) expected as a function of the stage in the task.

\bibliography{references}
\bibliographystyle{IEEEtran}

\end{document}